\documentclass[twocolumn]{elsart}
\usepackage{rotating,epsfig,picinpar,wrapfig,floatflt,graphicx}

\usepackage{amssymb}

\begin{document}
\newcommand{\An}      {\mbox{$ {\mathrm A}^0                               $}}
\newcommand{\nm}      {\mbox{$ \nu_m                                   $}}
\newcommand{\mnu}     {\mbox{$ m_{\nu}                                 $}}
\newcommand{\mnm}     {\mbox{$ m_{\nu_m}                               $}}
\newcommand{\hn}      {\mbox{$ {\mathrm h}^0                               $}}
\newcommand{\Zn}      {\mbox{$ {\mathrm Z}^0                               $}}
\newcommand{\Hn}      {\mbox{$ {\mathrm H}^0                               $}}
\newcommand{\HP}      {\mbox{$ {\mathrm H}^+                               $}}
\newcommand{\HM}      {\mbox{$ {\mathrm H}^-                               $}}
\newcommand{\CC}      {\mbox{$ {\mathrm CC}                                $}}
\newcommand{\NC}      {\mbox{$ {\mathrm NC}                                $}}
\newcommand{\Wp}      {\mbox{$ {\mathrm W}^+                               $}}
\newcommand{\Wpm}     {\mbox{$ {\mathrm W}^{\pm}                           $}}
\newcommand{\Wm}      {\mbox{$ {\mathrm W}^-                               $}}
\newcommand{\WW}      {\mbox{$ {\mathrm W}^+{\mathrm W}^-                  $}}
\newcommand{\ZZ}      {\mbox{$ {\mathrm Z}^0{\mathrm Z}^0                  $}}
\newcommand{\HZ}      {\mbox{$ {\mathrm H}^0 {\mathrm Z}^0                 $}}
\newcommand{\GW}      {\mbox{$ \Gamma_{\mathrm W}                          $}}
\newcommand{\Zg}      {\mbox{$ \Zn \gamma                                  $}}
\newcommand{\sqs}     {\mbox{$ \sqrt{s}                                    $}}
\newcommand{\ee}      {\mbox{$ {\mathrm e}^+ {\mathrm e}^-                 $}}
\newcommand{\eeWW}    {\mbox{$ \ee \rightarrow \WW                         $}}
\newcommand{\MeV}     {\mbox{$ {\mathrm{MeV}}                              $}}
\newcommand{\MeVc}    {\mbox{$ {\mathrm{MeV}}/c                            $}}
\newcommand{\MeVcc}   {\mbox{$ {\mathrm{MeV}}/c^2                          $}}
\newcommand{\GeV}     {\mbox{$ {\mathrm{GeV}}                              $}}
\newcommand{\GeVc}    {\mbox{$ {\mathrm{GeV}}/c                            $}}
\newcommand{\GeVcc}   {\mbox{$ {\mathrm{GeV}}/c^2                          $}}
\newcommand{\TeV}     {\mbox{$ {\mathrm{TeV}}                              $}}
\newcommand{\TeVc}    {\mbox{$ {\mathrm{TeV}}/c                            $}}
\newcommand{\TeVcc}   {\mbox{$ {\mathrm{TeV}}/c^2                          $}}
\newcommand{\MZ}      {\mbox{$ m_{{\mathrm Z}^0}                           $}}
\newcommand{\MW}      {\mbox{$ m_{\mathrm W}                               $}}
\newcommand{\MA}      {\mbox{$ m_{\mathrm A}                               $}}
\newcommand{\GF}      {\mbox{$ {\mathrm G}_{\mathrm F}                     $}}
\newcommand{\MH}      {\mbox{$ m_{{\mathrm H}^0}                           $}}
\newcommand{\MHP}     {\mbox{$ m_{{\mathrm H}^\pm}                         $}}
\newcommand{\MSH}     {\mbox{$ m_{{\mathrm h}^0}                           $}}
\newcommand{\MT}      {\mbox{$ m_{\mathrm t}                               $}}
\newcommand{\GZ}      {\mbox{$ \Gamma_{{\mathrm Z}^0}                      $}}
\newcommand{\TT}      {\mbox{$ \mathrm T                                   $}}
\newcommand{\UU}      {\mbox{$ \mathrm U                                   $}}
\newcommand{\alphmz}  {\mbox{$ \alpha (m_{{\mathrm Z}^0})                  $}}
\newcommand{\alphas}  {\mbox{$ \alpha_{\mathrm s}                          $}}
\newcommand{\alphmsb} {\mbox{$ \alphas (m_{\mathrm Z})
                               _{\overline{\mathrm{MS}}}                   $}}
\newcommand{\alphbar} {\mbox{$ \overline{\alpha}_{\mathrm s}               $}}
\newcommand{\Ptau}    {\mbox{$ P_{\tau}                                    $}}
\newcommand{\mean}[1] {\mbox{$ \left\langle #1 \right\rangle               $}}
\newcommand{\mydeg}   {\mbox{$ ^\circ                                      $}}
\newcommand{\qqg}     {\mbox{$ {\mathrm q}\bar{\mathrm q}\gamma            $}}
\newcommand{\Wev}     {\mbox{$ {\mathrm{W e}} \nu_{\mathrm e}              $}}
\newcommand{\Zvv}     {\mbox{$ \Zn \nu \bar{\nu}                           $}}
\newcommand{\Zee}     {\mbox{$ \Zn \ee                                     $}}
\newcommand{\ctw}     {\mbox{$ \cos\theta_{\mathrm W}                      $}}
\newcommand{\thw}     {\mbox{$ \theta_{\mathrm W}                          $}}
\newcommand{\thetabar}{\mbox{$ \theta^*                                    $}}
\newcommand{\phibar}  {\mbox{$ \phi^*                                      $}}
\newcommand{\thetapl} {\mbox{$ \theta_+                                    $}}
\newcommand{\phipl}   {\mbox{$ \phi_+                                      $}}
\newcommand{\thetamin}{\mbox{$ \theta_-                                    $}}
\newcommand{\phimin}  {\mbox{$ \phi_-                                      $}}
\newcommand{\ds}      {\mbox{$ {\mathrm d} \sigma                          $}}
\newcommand{\jjlv}    {\mbox{$ j j \ell \nu                                $}}
\newcommand{\jjjj}    {\mbox{$ j j j j                                     $}}
\newcommand{\jjvv}    {\mbox{$ j j \nu \bar{\nu}                           $}}
\newcommand{\qqvv}    {\mbox{$ \mathrm{q \bar{q}} \nu \bar{\nu}            $}}
\newcommand{\qqll}    {\mbox{$ \mathrm{q \bar{q}} \ell \bar{\ell}          $}}
\newcommand{\jjll}    {\mbox{$ j j \ell \bar{\ell}                         $}}
\newcommand{\lvlv}    {\mbox{$ \ell \nu \ell \nu                           $}}
\newcommand{\dz}      {\mbox{$ \delta g_{\mathrm{W W Z}    }               $}}
\newcommand{\ptr}     {\mbox{$ p_{\perp}                                   $}}
\newcommand{\ptrjet}  {\mbox{$ p_{\perp {\mathrm{jet}}}                    $}}
\newcommand{\Wvis}    {\mbox{$ {\mathrm W}_{\mathrm{vis}}                  $}}
\newcommand{\gamgam}  {\mbox{$ \gamma \gamma                               $}}
\newcommand{\qaqb}    {\mbox{$ {\mathrm q}_1 \bar{\mathrm q}_2             $}}
\newcommand{\qcqd}    {\mbox{$ {\mathrm q}_3 \bar{\mathrm q}_4             $}}
\newcommand{\bbbar}   {\mbox{$ {\mathrm b}\bar{\mathrm b}                  $}}
\newcommand{\ffbar}   {\mbox{$ {\mathrm f}\bar{\mathrm f}                  $}}
\newcommand{\qqbar}   {\mbox{$ {\mathrm q}\bar{\mathrm q}                  $}}
\newcommand{\djoin}   {\mbox{$ d_{\mathrm{join}}                           $}}
\newcommand{\mErad}   {\mbox{$ \left\langle E_{\mathrm{rad}} \right\rangle $}}
\newcommand{\ie}     {{\it i.e. }}
\newcommand{\eg}     {{\it e.g. }}
\newcommand{\dumand}     {\sc Dumand}
\newcommand{\baikal}     {\sc Baikal}
\newcommand{\nestor}     {\sc Nestor}
\newcommand{\amanda}     {{\sc Amanda }}
\newcommand{\asta}     {{\sc Asta}}
\newcommand{\spase}     {{\sc Spase}}
\newcommand{\antares}    {\sc Antares}
\newcommand{\lambdaa}    {\mbox{$ \lambda_a                                $}}
\newcommand{\lambdas}    {\mbox{$ \lambda_s                                $}}
\newcommand{\lambdaeff}  {\mbox{$ \lambda_{eff}                            $}}
\newcommand{\lambdabub}  {\mbox{$ \lambda_{bub}                            $}}
\newcommand{\dimpact}    {\mbox{$  d_{impact}                              $}}
\newcommand{\likelihood} {\mbox{$ \mathcal{L}                              $} } 
\newcommand{\varphiom}   {\mbox{$\varphi_{\scriptsize OM}                     $}} 
\newcommand{\thetaom}    {\mbox{$\theta_{\scriptsize OM }                  $}} 
\newcommand{\numu}       {\mbox{$\nu_{\mu}                  $}} 
\newcommand{\nue}        {\mbox{$\nu_{e}                  $}} 
\newcommand{\nutau}      {\mbox{$\nu_{\tau}                  $}} 
\newcommand{\antinumu}       {\mbox{${\bar{\nu}}_{\mu}                  $}} 
\newcommand{\antinue}        {\mbox{${\bar{\nu}}_{e}                  $}} 
\newcommand{\antinutau}      {\mbox{${\bar{\nu}}_{\tau}                  $}} 
\def\leqsim{\mathbin{\;\raise1pt\hbox{$<$}\kern-8pt\lower3pt\hbox{\small$\sim$}\;}}
\def\geqsim{\mathbin{\;\raise1pt\hbox{$>$}\kern-8pt\lower3pt\hbox{\small$\sim$}\;}}
 \newcommand{\stot}      {\mbox{$S_\mathrm{tot} $}} 
 \newcommand{\deltamu}     {\mbox{ $\Delta\mu $}} 
 \newcommand{\ma}      {\mbox{$MA $}} 
 \newcommand{\masym}      {\mbox{$MA_{sym} $}} 
 \newcommand{\res}      {\mbox{$RES $}} 
 \newcommand{\tw}      {\mbox{$n_{\mathrm{window}} $}} 
\newcommand{\umlaut}{\"} 
\newcommand{\eprint}{\textsf} 

\newcommand{\journalfont}{\rm}  
\newcommand{\jou}[1]{{\journalfont #1\ }}
\newcommand{\joudef}[2]{\newcommand #1{\jou{\ignorespaces #2}}}

\joudef{\aaa}    { Astron.\ Astrophys.}
\joudef{\aip}    { Adv.\ Phys.}
\joudef{\adm}    { Adv.\ Math.}
\joudef{\am}     { Ann.\ Math.}
\joudef{\apny}   { Ann.\ Phys.\ (N.Y.)}
\joudef{\apj}    { Astrophys.\ J.}
\joudef{\apjs}   { Astrophys.\ J.\ Suppl.}
\joudef{\app}    { Astropart.\ Phys.}
\joudef{\cjp}    { Can.\ J.\ Phys.}
\joudef{\cmp}    { Commun.\ Math.\ Phys.}
\joudef{\cqg}    { Class.\ Quantum Grav.}
\joudef{\eul}    { Europhys. \Lett.}
\joudef{\faa}    { Funct.\ Anal.\ Appl.}
\joudef{\grg}    { Gen.\ Rel.\ Grav.}
\joudef{\ijmpd}  { Int.\ J.\ Mod.\ Phys.\ D}
\joudef{\ijtp}   { Int.\ J.\ Theor.\ Phys.}
\joudef{\invm}   { Invent.\ Math.}
\joudef{\jm}     { J.\ Math.}
\joudef{\jmp}    { J.\ Math.\ Phys.}
\joudef{\jpa}    { J.\ Phys.\ A}
\joudef{\mnras}  { Mon.\ Not.\ R.\ Ast.\ Soc.}
\joudef{\mpla}   { Mod.\ Phys.\ Lett.\ A} 
\joudef{\nature} { Nature}
\joudef{\nc}     { Nuovo Cim.}
\joudef{\nim}    { Nucl.\ Instrum.\ Meth.}
\joudef{\np}    { Nuc.\ Phys.}
\joudef{\npb}    { Nucl.\ Phys.\ B}
\joudef{\npbsuppl}    { Nucl.\ Phys.\ B\ Suppl.}
\joudef{\ph}     { Physica}
\joudef{\pla}    { Phys.\ Lett. A}
\joudef{\plb}    { Phys.\ Lett. B}
\joudef{\pr}     { Phys.\ Rev.}
\joudef{\prd}    { Phys.\ Rev.\ D}
\joudef{\prep}   { Phys.\ Rep.}
\joudef{\prl}    { Phys.\ Rev.\ Lett.}
\joudef{\prsla}  { Proc.\ Roy.\ Soc.\ Lond.\ A}
\joudef{\ptp}    { Prog.\ Theor.\ Phys.}
\joudef{\ptps}   { Prog.\ Theor.\ Phys.\ Suppl.}
\joudef\rmp      { Rev.\ Mod.\ Phys.}
\joudef\spj      { Sov.\ Phys.\ JETP}
\joudef\zpc      { Z.\ Phys.\ C}
\newcommand{\Ecms}    {\mbox{$ E_{\mathrm{cms}}                            $}}
\newcommand{\Evis}    {\mbox{$ E_{\mathrm{vis}}                            $}}
\newcommand{\Erad}    {\mbox{$ E_{\mathrm{rad}}                            $}}
\newcommand{\Mvis}    {\mbox{$ M_{\mathrm{vis}}                            $}}
\newcommand{\Mevt}    {\mbox{$ M_{\mathrm{evt}}                            $}}
\newcommand{\pvis}    {\mbox{$ p_{\mathrm{vis}}                            $}}
\newcommand{\Minv}    {\mbox{$ M_{\mathrm{inv}}                            $}}
\newcommand{\Mhfit}{\; \hat{m}_{H^0} }
\def\longrightbracearrow{\raise0.5em\hbox{$|$}\mkern-9.0mu\longrightarrow}
\let\lrbracearrow=\longrightbracearrow   

\newcommand{\Zto}   {\mbox{$\mathrm Z^0 \to$}}
\newcommand{\SM}   {\mbox{\rm Standard Model }}

\def\NPB#1#2#3{{\it Nucl.~Phys.} {\bf{B#1}} (19#2) #3}
\def\PLB#1#2#3{{\it Phys.~Lett.} {\bf{B#1}} (19#2) #3}
\def\PRD#1#2#3{{\it Phys.~Rev.} {\bf{D#1}} (19#2) #3}
\def\PRL#1#2#3{{\it Phys.~Rev.~Lett.} {\bf{#1}} (19#2) #3}
\def\ZPC#1#2#3{{\it Z.~Phys.} {\bf C#1} (19#2) #3}
\def\PTP#1#2#3{{\it Prog.~Theor.~Phys.} {\bf#1}  (19#2) #3}
\def\MPL#1#2#3{{\it Mod.~Phys.~Lett.} {\bf#1} (19#2) #3}
\def\PR#1#2#3{{\it Phys.~Rep.} {\bf#1} (19#2) #3}
\def\RMP#1#2#3{{\it Rev.~Mod.~Phys.} {\bf#1} (19#2) #3}
\def\HPA#1#2#3{{\it Helv.~Phys.~Acta} {\bf#1} (19#2) #3}
\def\NIMA#1#2#3{{\it Nucl.~Instr.~and~Meth.} {\bf#1} (19#2) #3} 

\begin{frontmatter}

\title{Search for Supernova Neutrino-Bursts with the \amanda Detector}
\author[label9]{J.~Ahrens}, 
\author[label1]{X.~Bai}, 
\author[label11]{G.~Barouch}, 
\author[label8]{S.W.~Barwick}, 
\author[label7]{R.C.~Bay}, 
\author[label9]{T.~Becka}, 
\author[label2]{K.-H.~Becker}, 
\author[label3]{D.~Bertrand}, 
\author[label4]{A.~Biron}, 
\author[label8]{J.~Booth}, 
% ...

  \address[label1] {Bartol Research Institute, University of Delaware, Newark, DE 19716, USA}
  \address[label2] {Fachbereich 8 Physik, BUGH Wuppertal, D-42097 Wuppertal, Germany}
  \address[label3] {Brussels Free University, Science Faculty CP230, Boulevard du Triomphe, B-1050 Brussels, Belgium}
  \address[label4] {DESY-Zeuthen, D-15735 Zeuthen, Germany}
  \address[label5] {Dept. of Technology, Kalmar University, S-39182 Kalmar, Sweden}
\clearpage
\newpage
  \address[label6] {Lawrence Berkeley National Laboratory, Berkeley, CA 94720, USA}
  \address[label7] {Dept. of Physics, University of California, Berkeley, CA 94720, USA}
  \address[label8] {Dept. of Physics and Astronomy, University of California, Irvine, CA 92697, USA}
  \address[label9] {Institute of Physics, University of Mainz, Staudinger Weg 7, D-55099 Mainz, Germany}
  \address[label10] {Dept. of Physics and Astronomy, University of Pennsylvania, Philadelphia, PA 19104, USA}
  \address[label11] {Dept. of Physics, University of Wisconsin, Madison, WI 53706, USA}
  \address[label12] {Dept. of Radiation Sciences, Uppsala University, S-75121 Uppsala, Sweden}
  \address[label13] {Fysikum, Stockholm University, S-11385 Stockholm, Sweden}

\begin{abstract}
The core collapse of a massive star in the Milky Way will produce a neutrino burst, 
intense enough to be detected by existing underground detectors. The AMANDA neutrino 
telescope located deep in the South Pole ice can detect MeV neutrinos by a collective
rate increase in all photo-multipliers on top of dark noise.  The main source 
of light comes from positrons produced in the CC-reaction of anti-electron neutrinos on free 
protons $\antinue + p \rightarrow e^+ + n$. This paper describes the first supernova 
search performed on the full sets of data taken during 1997 and 1998 (215 days of live
time) with 302 of the detector's optical modules. No candidate events resulted from
this search. The performance of the detector is calculated, yielding a 70$\%$ 
coverage of the Galaxy with one background fake per year with 90$\%$ 
efficiency for the  detector configuration under study.
An upper limit  at the 90$\%$ c.l. on the rate of stellar collapses in the Milky Way 
is derived, yielding 4.3 events per year. A trigger algorithm is presented and its
performance estimated. Possible improvements of the detector hardware are reviewed.
\end{abstract}


\end{frontmatter}

\section{Introduction}
\label{sec:intro}
Astronomical observations have customarily been carried out by detecting photons, 
from radio-waves to gamma rays, with each new range of energy uncovered leading to fresh
discoveries.
Over the years, the electro-magnetic spectrum has been well covered, leaving no
new frequency-gaps to explore. \par
Neutrino astronomy is still in its infancy, but it is a field which has been growing
over the past few decades. Besides providing a new probe to investigate the universe,
the neutrino is also fundamentally different from the photon in that it interacts
only weakly and can thus still be observed after passing through large
amounts of matter. 
This is a mixed blessing, however, since the detection of neutrinos on Earth is made
more difficult by the very same reason and requires correspondingly large detector
volumes. Furthermore, neutrinos remain undeflected by magnetic fields. \par
Until now, the only extra-terrestrial sources of neutrinos that have been observed are the 
Sun and the supernova SN1987A. In both cases, the energies seen were at or below a few tens
of MeV. The \antinue~-burst emitted in the 1987 supernova event was detected simultaneously by 
the IMB~\cite{IMB} and Kamiokande~II~\cite{k2,hirata88} water detectors, a few hours ahead of its 
optical counterpart. Due to the distant location of the supernova in the Large Magellanic 
Cloud ($\sim$52 kpc from us), only 20 neutrinos were collected in total. This was still enough to 
confirm that most of the energy was released in the form of neutrinos and also to validate the 
essential predictions of models describing the mechanism of gravitational collapse supernovae. 
SN1987A proved also that relevant particle 
physics information can be extracted from astrophysical events. 
The data was used to set an upper limit on the mass of the \antinue, 
its lifetime, its magnetic moment and the number of leptonic 
flavours.\par
However, although the rough features in terms of energies, duration and flux were supported by 
the observations, they said next to nothing about the details of the burst
development. For this, hundreds 
if not thousands of neutrinos would be needed. Furthermore, both experiments were insensitive 
to flavors other than \antinue, which are expected to carry away most of the supernova energy.
The detection of a type II supernova in the Milky Way would provide us with a unique
opportunity to study the details of the gravitational collapse 
of a star. Several low-energy neutrino detectors exist already
(LVD, Super-Kamiokande, SNO \cite{scholberg3} and Baksan \cite{baksan}) which
will be able to shed light on the core collapse of a massive star, should such an event
occur during their lifetime. \par
\amanda (Antarctic Muon and Neutrino Detector Array~\cite{b4}) is one of several high energy 
neutrino telescopes. Another existing detector is NT200 in lake Baikal \cite{baikal99}, 
whereas others are under development (ANTARES \cite{antares2000}, NESTOR \cite{nestor2000} 
and NEMO \cite{nemo_icrc99}).
\amanda consists of optical modules (OMs) buried 1500-2000 m deep in the
Antarctic ice sheet. Each OM is made up of a photo-multiplier tube (PMT) enclosed in a
pressure-resistant glass vessel and connected to the surface electronics by an
electrical cable supplying power and transmitting the PMT signals. 
\par
\amanda is designed for the observation 
of TeV neutrino-sources, utilizing the large volume of transparent glacier ice available at 
the South Pole as a Cherenkov medium. In spite of the much lower neutrino energies of
${\mathcal{O}}$(10 MeV) involved in  a burst, it has been shown \cite{sn_halzen} that a
detector of this type could also be used successfully to monitor our Galaxy for
supernova events (the description of a method for neutrino telescopes using ocean water can be 
found in \cite{pryor}).
Since the cross-section for inverse $\beta$ decay reaction on protons $\antinue + p
\rightarrow n + e^+ $ in the ice exceeds the cross-sections for the other neutrino
flavors and targets, $\antinue$ events are dominating 
\footnote{Note that the scattering of \antinue ~on $^{16}$O is negligible at the 
expected energies.}.
During the estimated $\sim$ 10 sec duration of a neutrino-burst, the Cherenkov light
produced by the positron tracks will increase the counting rate of
all the PMTs in the detector above their average value. This
effect, when considered as a collective behavior, could be seen clearly even if the
increase in {\it each} PMT would not be statistically significant. An observation made
over a time window of several sec could therefore provide a detection of a 
supernova, before its optical counterpart is observed. 
The stable and low background noise in \amanda (absence of $^{40}K$ and of bioluminescence 
in the ice) is a clear asset for this method. 
\par
 In this paper, the 302 OMs of the \amanda B10 stage completed in 1997 are
 used for the analysis. In the next section, we summarize the theoretical
expectations for the neutrino burst which is emitted when
a massive star undergoes gravitational core collapse. An account is then given of the
detector hardware, as well
as of the data collected with it. The analysis method is explained next and the results 
obtained from its application to experimental data are presented. A study
of the detector performance follows. Next, the principle of an
online trigger algorithm is described. We end with a discussion of possible
improvements of the detector.
Early searches for supernova neutrino-burst signals with the \amanda detector have
been presented in \cite{sn_rome,desy_proposal,icrc99}.

\section{Theoretical preliminaries}
\label{sec:theory}
When the core of a massive star ($M \ge 8 M_\odot $) runs out of
nuclear fuel, it collapses and ejects the outer mantle in a SN explosion of type II/Ib/Ic 
~\cite{kennicutt}. Only $\sim 1\%$ of the energy is released in
kinetic and optical form, whereas the
remaining $99\%$ of the gravitational binding energy change, about $3\times 10^{53}$ ergs,
is carried away by neutrinos~\cite{burrows}. \par
During the process, the inner core reaches nuclear densities, bringing the collapse to an abrupt
stop, and a shock-wave is formed. The front of this shock is driven outward, through
the in-falling  material, passing through the
neutrinosphere (the location where the material changes from opaque to transparent to neutrinos) 
on its way, until a point where it stalls . \par
During the first 10 msec, a \nue~burst from the neutronization process $e^- + p
\rightarrow n + \nue$ releases $\sim 10^{52}$ ergs.
Although neutrinos interact only weakly, the densities built up during the collapse 
are so high that they cannot stream out. Instead, they are trapped and
diffuse out over a time scale of several sec. When they finally reach
the neutrinosphere they can escape,  with a 
thermal  spectrum which is approximately Fermi-Dirac~\cite{burrows}.
The trapped neutrinos are produced in pairs through the reaction $e^- + e^+
\stackrel{Z}{\rightarrow}  \nu +
\bar{\nu}$ for all lepton flavors. In addition, the $\nue \antinue$ pairs are also
produced via $p(n) + e^{-(+)} \stackrel{W}{\rightarrow} n (p) + \stackrel{(-)}{\nu_e}$. 
They are produced with distinctive energies, because the neutrino-spheres for each type
are located at depths with different temperatures.
The $\numu$ and $\nutau$ and their anti-particles have a mean energy of  
$\left < E \right > \approx 25 \rm \,MeV$. The $\antinue$ have a mean energy
$\left < E \right > \approx 16 \rm \, MeV$ and the corresponding value for $\nue$
neutrinos is $\left < E \right > \approx 11 \rm \,MeV$~\cite{raffelt}. 
In total, the radiated energy is expected to be equally distributed over each flavor of
neutrinos and anti-neutrinos \cite{beacomsuperk}. The luminosity-profile can be
described by a quick rise over a few msec and more and then falling over a
time of $\mathcal{O}({\mbox{several sec}})$, roughly like an exponential with a
time constant $\tau= 3  ~\rm sec$~\cite{beacom}. During the late phase the neutrino 
luminosity possibly follows a power law with index $\sim 1 \pm 0.5$~\cite{burrows}.
The detailed form of the neutrino luminosity used below is less 
important than the general shape features and their characteristic
durations~\cite{beacom}. 
\begin{figure}[htp]
 \centering
  \includegraphics*[width=0.45\textwidth]{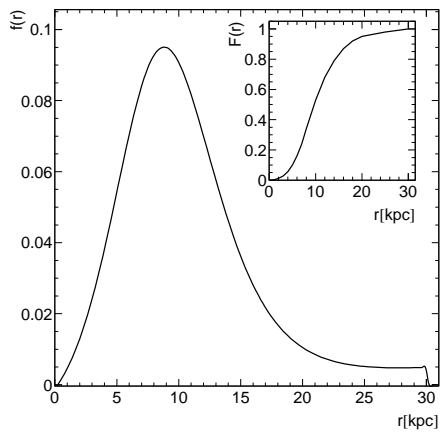}
        \caption{
\footnotesize{The distribution $f(r)$ of progenitor stars in the Milky Way, located 
within a distance  $r$ from the Earth. Inset: the corresponding cumulative distribution $F(r)$
(taken from a model by Bahcall \& Soneira~\cite{soneira,piran}).
The steep behavior around $\sim$ 8 kpc is due to the high star density in the Galactic bulge.
}}
        \label{fig:fraction}
\end{figure}
\par
Currently, the rate of galactic stellar collapses is
estimated to be one every 40 years~\cite{sn_rate}, with a best
experimental upper limit of 1 every 5 years~\cite{baksan}. At least three 
underground neutrino detectors (Super-Kamiokande~\cite{superkpro}, Baksan~\cite{baksan}, LVD~\cite{lvd}) 
have a sensitivity high enough to cover the Galaxy and beyond ( the distribution of
progenitor stars in the Milky Way is shown in Fig.\ref{fig:fraction}).
\section{Description of the detector}
\label{sec:detector}
\begin{figure}[htp]
 \centering
  \includegraphics*[width=0.5\textwidth]{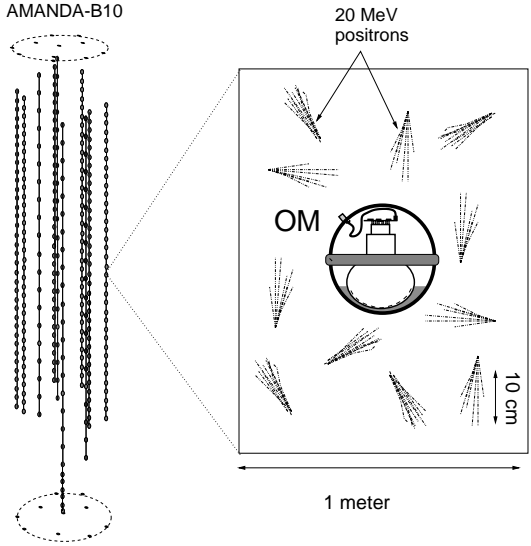}
           \caption{
\footnotesize{
View of the detector and of the expected positron tracks
produced in CC  $\bar{\nu_{e}}$ interactions within {\sc Amanda} }}
           \label{fig:nu-e}
\end{figure}

 The \amanda detector has been deployed over several years\cite{b4}. Each deployment phase
consists of drilling holes in the ice and lowering strings of OMs connected to
electrical cables (see Fig.\ref{fig:nu-e}).
In this text, we will limit ourselves to the study of data taken in 1997 and 1998, with
ten strings (302  OMs) in operation during these two years.
\par The first four strings were deployed in 1995-1996, using coaxial cable. Strings 5-10 followed in
1996-1997, using twisted-pair cable. Hamamatsu photo-multipliers (PMTs) of type R5912-02 
were used in all OMs. The PMTs of strings 1-4 are enclosed in spheres  manufactured by 
Billings, whereas the remaining strings use spheres with better transparency
\cite{desy_proposal}, produced by Benthos company. 
The AMANDA-supernova detector was part of the AMANDA-SNMP
(SuperNova-MonoPole) data-acquisition system (DAQ) \cite{desy_proposal}, which operates
independently from the AMANDA muon-trigger and DAQ. It consisted of 
custom-made CAMAC modules and is read out by a Macintosh via a
standard SCSI interface.\par
The idea is to continuously measure the counting rates of all OM channels separately 
and store that data for further analysis. For this, each channel counted the arriving
OM-pulses synchronously into a separate 12-bit counter during a fixed time interval
of 500 msec which is synchronized by a GPS-clock.\par
Correlated noise following OM-pulses was suppressed by a common programmable 
dead-time in the range of 100 ns to 12.8 $\mu$s (which was set to 10 $\mu$s). \par
The present detector configuration has nine additional strings, bringing the total number of OMs to 
680, and the data is currently taken with an upgraded VME/Linux based DAQ.
\subsection{Low-energy \antinue ~in \amanda}
\label{sec:lownue}
\begin{figure}[htp]
 \centering
  \includegraphics*[width=0.5\textwidth]{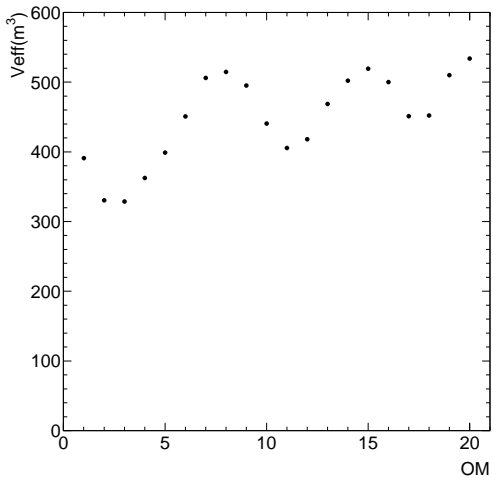}
           \caption{
\footnotesize{
Effective volume as a function of OM number in string 1. The variations are due to the change in ice properties as a function of depth. The OMs are numbered from their highest to lowest location.}}
           \label{fig:veff}
\end{figure}

The anti-electron neutrinos produced in a supernova follow a Fermi-Dirac distribution
with $\left < E_{\antinue} \right > \approx 16 \rm \: MeV$
, which leads to a peak value of the positron
energy distribution $E_{\mathrm{e^+}}^{\mathrm{peak}} \sim $ 20 MeV~\cite{sn_halzen}.  The
corresponding track length is thus $\sim$12 cm \cite{halzen92,halzennew,jacobsen}. 
The expected signal for a SN1987A-type supernova has been simulated~\cite{halzennew,jacobsen}, 
yielding a predicted excess in the number of photo-electrons per PMT: 
\begin{eqnarray} \label{eq:halzen_signal}  N_{\mathrm{p.e.}}\sim 11 \cdot \left [\frac{\rho_{\mathrm{ice}} \cdot V_{\mathrm {eff}}}{M_{\mathrm{Kam}}}\right
] \left [\frac{52~{\rm kpc}}{d_{{\rm kpc}}} \right ] ^2\end{eqnarray} 
 where  11 is the number of events detected by Kamiokande II, ${M_{\mathrm{Kam}}}$ is the 
effective mass of Kamiokande II and ${d_{{\rm kpc}}}$ is the
distance between the supernova and Earth. The value of ${M_{\mathrm{Kam}}}$ is 2.14 kton, 
times a factor 0.8 to correct for the fact that Kamiokande II had an energy threshold,
whereas the \amanda supernova system does not (see \cite{jacobsen}).
The ice density, $\rho_{\mathrm{ice}}$, is 0.924 $\mathrm{g/cm^3}$.
In preliminary studies, the effective volume $V_{\mathrm {eff}}$ of one OM was
found to be approximately proportional to the absorption length of ice
and estimated to be $414~\rm{m^3}$ (using an ice-model with 90 m absorption
length)~\cite{jacobsen}. A subsequent detailed treatment was made, based on the
depth-dependent ice properties described in \cite{icrc-kurt}. This showed that
$V_{\mathrm {eff}}$ varies, depending
on the OM location. The resulting values for string 1 as a function of OM
number are shown in Fig.\ref{fig:veff}.
The different absorption of Benthos and Billings glass spheres has also been taken
into account.
Following Eq.\ref{eq:halzen_signal}, we expect $\sim$100 counts/OM during 10
sec~\cite{sn_halzen} for a SN1987A-like supernova at 8 kpc distance from the Earth. 
\hfill \par
The summed rate over $N_{\mathrm {OM}}$ optical modules in the presence of a supernova is:
\begin{eqnarray} \label{eq:halzen_bg} 
\lefteqn{S_\mathrm{tot} = S_{\mathrm{noise}} +S_{\mathrm {signal}} =}\\\nonumber
 & &\hspace{2 cm} \sum_{i=1}^{N_{\mathrm {OM}}}(R^{\mathrm{noise}}_i+R^{\mathrm {signal}}_i)  \end{eqnarray}  
$S_{\mathrm{noise}}$ is the dark noise summed over all OMs and $S_{\mathrm {signal}}$ is the summed
rate of photo-electrons resulting from \antinue-interactions. 
$R^{\mathrm{noise}}_i$ and $R^{\mathrm {signal}}_i$ are the rates of 
background and signal counts, respectively, for each
individual OM. If a supernova would occur, one should observe a sudden increase in
$S_\mathrm{tot}$, from $S_{\mathrm{noise}}$ to $S_{\mathrm{noise}} +S_{\mathrm {signal}}$~\cite{sn_halzen}.
 \subsection{The noise background}
Thanks to the low temperature in the ice, $\sim-35^\circ \mathrm C$, the PMT noise is
strongly reduced. Furthermore, the ice is a low-noise medium, free from natural
radioactivity and bioluminescence.
However, in order to detect a small excess due to the signal, it is not the level of the
dark-noise $S_{\mathrm{noise}}$, as much as its
fluctuation that has to be kept low.
If the dark noise from the OMs would be purely Poissonian, the
fluctuation of $S_{\mathrm{noise}}$~during $\Delta t=10$ sec would be: \begin{eqnarray} 
\label{eq:halzen_sigma} \sigma_{S_{\mathrm{noise}}}=\sqrt{\Delta t \cdot
R_{\mathrm{noise}}\cdot N_{\mathrm {OM}}} \end{eqnarray} where $ R_{\mathrm{noise}}$ is the typical
OM noise rate in Hz. \par
In reality, the situation has been found to be more complex and the Poisson assumption
had to be modified for various reasons. Measurements in the laboratory have revealed two
types of after-pulses due to ionized rest gases, one delayed by $\sim$1.5 $\mu$s and the 
other by $\sim$6 $\mu$s. 
This situation has been improved, by implementing a 10
$\mu$s artificial dead-time to suppress after-pulsing \cite{desy_proposal}.\par
At that stage the OM dark-noise follows a Gaussian
distribution with average rates of 300 Hz for OMs in strings 1-4
and 1100 Hz for OMs in strings 5-10. 
The difference between the two groups is due to the higher
potassium content of the spheres used in the second set.
One of the processes by which
$\beta$-decaying $^{40}K$ increases the counting rates is by scintillation\footnote{Cherenkov light would
have a signature of high amplitude pulses, which are not observed.}, where the
primary electrons produce secondary photons.
 This non-Poissonian process has been studied and is described in~\cite{teich1}.
Estimates of the potassium content in the different spheres used were made in the
laboratory, yielding
  2$\%$ for Benthos and $\sim 0.1\%$ for Billings spheres. The potassium content of 
the PMT glass is only $0.013\%$; moreover, the PMT envelope is much
thinner than the glass vessel.
\begin{figure}[h]
  \centering
  \includegraphics*[width=0.45\textwidth]{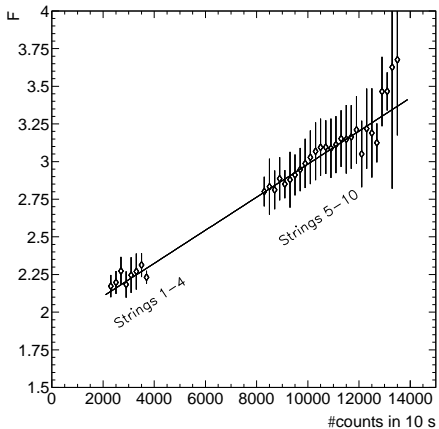}
           \caption{
\footnotesize{
 F as a function of the noise, showing the strong correlation
with the average of the measured rates. Note that the ordinate starts at $F=1.5$. 
}}
  \label{fig:stable_om}
  \end{figure}
\par It was observed that the variance of  the OM dark-noise is
larger than expected. A Fano factor can be used to describe this deviation from a
Poissonian behavior (see \cite{teich1}), $F=\sigma^2(N)/\langle N \rangle$, where $\langle {N}
\rangle$ is the average number of counts during a 10 sec intervals. For a Poisson
process, $F=1$. We get $F\approx 2.2$ for strings 1-4 and  $F\approx 3 $ for strings
5-10. This is shown in Fig.\ref{fig:stable_om}, where the behavior of $F$ as
a function of the dark noise is fitted with a line.
\par
Another part of the dark noise is due to Cherenkov light from atmospheric
muons passing nearby. Even though the muon flux is considerably attenuated by the ice-cover,
this contribution is not completely negligible. 
We have estimated a noise rate of 25 Hz due to muon light at the topmost OM and 12 Hz at the 
bottom.

\section{Data}
\label{sec:data}
The data were  rebinned before analysis, using a time window of 10 sec 
 for the rate calculation instead of the original 0.5 sec used in the DAQ. Henceforth, each 10
sec interval will be referred to as an {\it event} and all rates 
will be expressed in [Hz]. Each run has a typical length 14.6 hour.
\par
The quality of the data depends on two main factors: the stability of each 
OM over long-term periods and external disturbances of the DAQ which can render a 
run useless. 
 \begin{figure}[h]
   \centering
  \includegraphics*[width=0.4\textwidth]{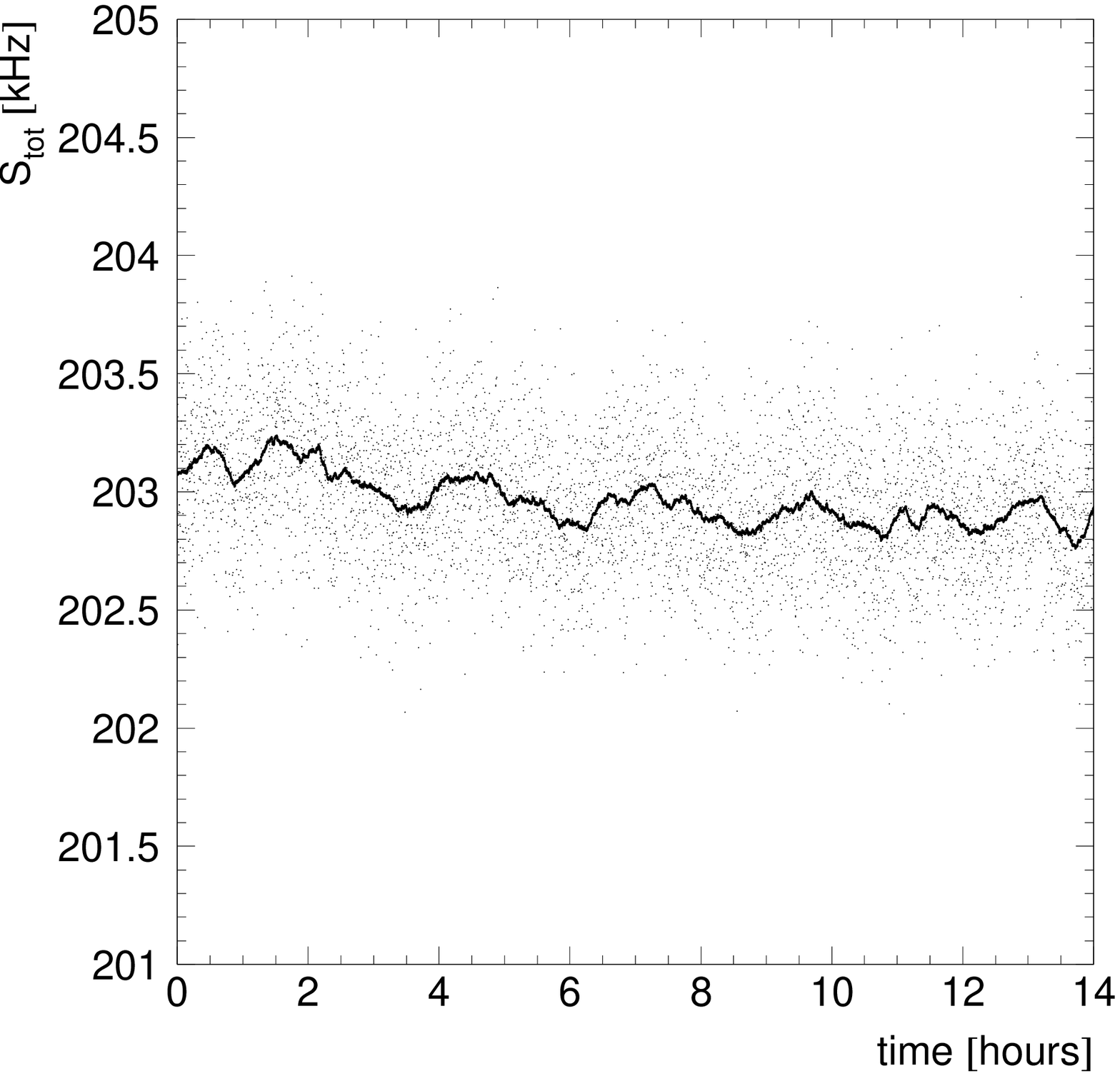}
  \includegraphics*[width=0.4\textwidth]{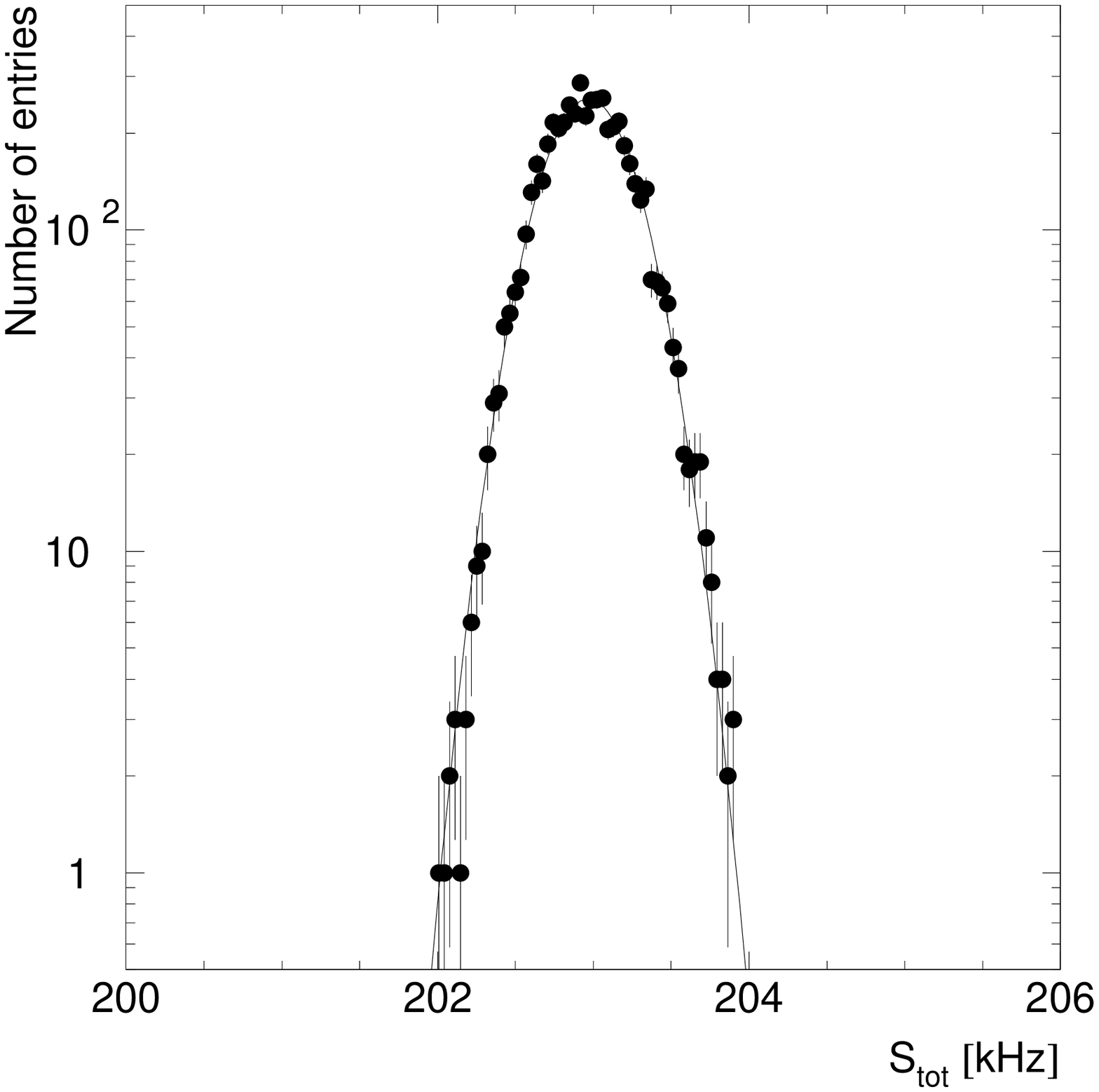}
         \caption{
 \footnotesize{Top: summed counting rate $S_\mathrm{tot}$ as a function
 of time for the selected OMs. The full line represents the moving average. Bottom: the
corresponding $S_\mathrm{tot}$-distribution (dots) fitted with a Gaussian (curve).
 }}
   \label{fig:stable_file}
  \end{figure}
Data-quality varied
considerably over the year for several OMs. In order to reach a state where a fixed subset of
OMs were stable in a fixed collection of runs, we developed an iterative
selection algorithm. As an input, we produced a binary table describing the quality of each
OM in each run, a 1 representing a `{\it good}' OM in the corresponding data-run and
a 0 representing a low quality OM in that run. An OM is meant to be `{\it good}' in a
run if it exhibits a Gaussian behaviour with a noise rate consistent with its average
over a year.
\par The resulting matrix was then processed iteratively according to the following steps:\\
1. The sum of each row and each column is calculated.\\
2. The OM {\it or} run corresponding to the smallest sum found in step (1) is removed
from the table. Here, the basic assumption is that OMs and runs are of equal value for
the analysis.\\
3. The iteration is stopped if there are no 0's left in the table,  {\it i.e.} when each
OM left is stable in each run left. Otherwise, the procedure is repeated by going back
to step (1).
\begin{table}[htp]
\begin{center}
\begin{tabular} {lc c c}
\hline\hline
1997  & Runs &Days&OMs  \\
Available  &            318 &   185&  302 \\
After cleaning &              179&    103 &   224 \\
\cline{1-4}   

 1998 & Runs &Days&OMs\\
Available  &   299 &   164&    302 \\
After cleaning  &   205&    112 &   235\\
\hline            \hline            
\end{tabular} 
\vfill{\ }
\caption{
\footnotesize{Summary of the  effect of data-cleaning, both for 1997 and 1998.}}
\label{tab:file_pmt}
\end{center}
\end{table}
Table~\ref{tab:file_pmt}  summarizes the effect that the cleaning has on the
live-time and number of selected OMs. The OM set is almost the same in
both 1997 and 1998 and the live-time difference is mainly due to calibration activity involving laser
light in 1997, which caused more runs to be rejected by the algorithm that year. 
Of the 224 OMs selected in 1997 and 235 in 1998, 52 come from the 80 OMs in
strings 1-4 and the remainder come from strings 5-10. 
\par 
The variation of ${S_\mathrm{tot}}$ is shown for one '{\it good}'  run in
Fig.\ref{fig:stable_file} (top). Possible deviations from a Gaussian distribution which are due to slow
drifts of the noise rates over time (see Fig.\ref{fig:stable_file}, bottom) can be corrected for, as described below.
\section{Analysis}
\label{sec:analysis}
\begin{figure}[h]
  \centering
  \includegraphics*[width=0.45\textwidth]{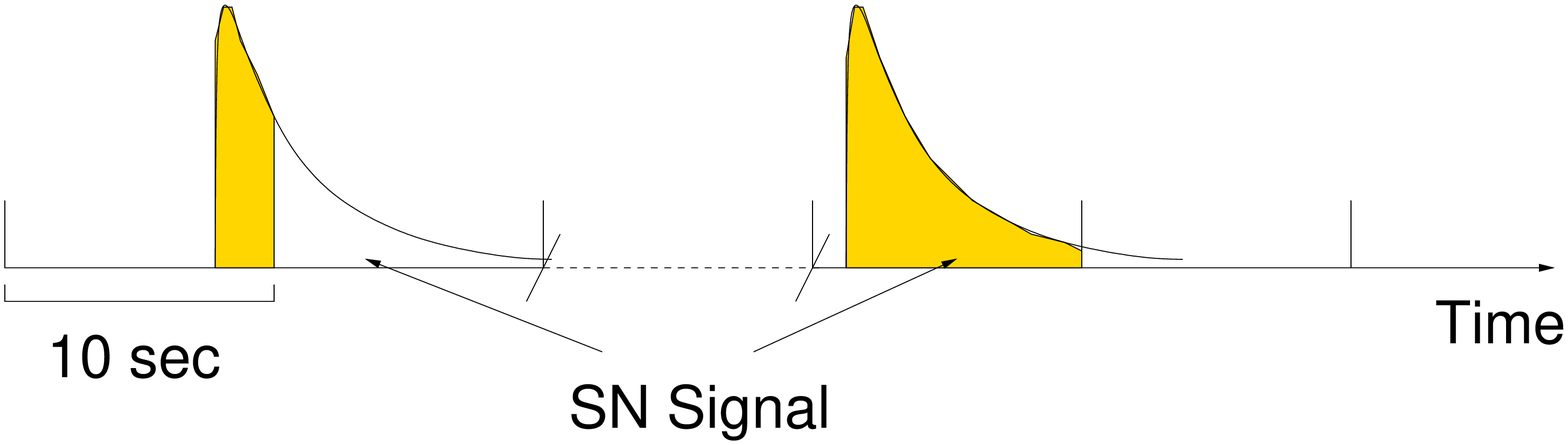}
        \caption{
\footnotesize{The location of the time window of  observation and 
the SN burst are uncorrelated,  resulting in a reduced sensitivity.
}}
  \label{fig:sn_onset}
\end{figure}

Though long enough to cover the supernova burst when synchronized with it, an
arbitrarily located time window of 10 sec will (on average) not match the onset of a
supernova signal (see Fig.\ref{fig:sn_onset}). 
If we consider the canonical supernova neutrino-burst time profile mentioned in
Sec.\ref{sec:intro}, the loss in signal due to the time-window location is
$\sim 20\%$. 
\par
The OM-stability is a crucial parameter for our purposes and it can be
influenced by several factors. Some of them, such as high-voltage variations, are of
known origin and their effect can be both understood and estimated. Long-time trends
are also present. {\it E.g.}, due to PMT aging, the dark noise decreases by $\sim 2\%$ per year.
These variations are beyond our control and can only be monitored. Other disturbances
are more irregular, such as human interventions on the front end electronics, or periods of 
severe weather conditions at the South Pole which have been observed to interfere with 
the data-taking.\par
 In order to remove monotonic trends and periodical fluctuations  on a scale of several hours 
and longer,  we subtract a moving average (\ma) from the \stot~values.
The residual deviation \res, defined by the difference between \stot~and \ma ~for each event ($i$), 
should then yield a time series stationary on scales of the order of an hour
and longer\cite{brockwell}:  \par
\begin{eqnarray}
\label{eq:moving_average}
\lefteqn{RES (S_\mathrm{tot}(i)) =}\nonumber \\ &&\hspace{1.7cm} S_\mathrm{tot}(i) - MA(S_\mathrm{tot}(i)) 
 \end{eqnarray} 
with the average taken over a time window of length ($\tw+1$) bins:
\begin{eqnarray}
\label{eq:moving_average}
\lefteqn{ MA(S_\mathrm{tot}(i)) = {\frac{1}{(\tw+1)}} \times} \nonumber \\ &&\hspace{1.7cm} \sum_{j=-\tw/2}^{\tw/2} S_\mathrm{tot}(i+j)  
 \end{eqnarray} 
The \ma -subtraction acts like a high-pass filter and was
checked not to affect our SN-burst
signal detection ability. \par
\begin{figure}[h]
  \centering
  \includegraphics*[width=0.4\textwidth]{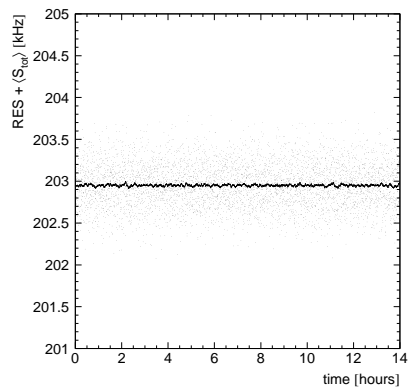}
        \caption{
\footnotesize{The same time series as in Fig.\ref{fig:stable_file}, after
moving-average subtraction and shifted by $\left < S_{tot} \right >$.
}
}
  \label{fig:sum_ma}
\end{figure}
\begin{figure}[htp]
 \centering
  \includegraphics*[width=0.5\textwidth]{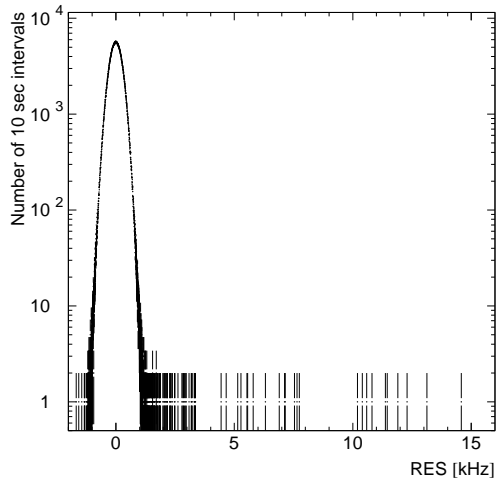}
        \caption{
\footnotesize{Distribution of residual summed noise \res ~for 215 days live-time of 1997
and 1998 data. A Gaussian fit of the peak of the distribution yields a standard
deviation $\sigma=265$ Hz.
}}
        \label{fig:only_ma}
\end{figure}

 For this off-line analysis, the sum is taken symmetrically over $(\tw+1)$
time-bins around the studied event and will therefore be referred to as $MA_\mathrm{sym}$.
We found that 1000 sec was an appropriate interval, correcting for most of the
variations observed in the data, as shown in Fig.\ref{fig:sum_ma}.
Fig.\ref{fig:only_ma}
shows the residuals of the summed noise, histogrammed for the 215 days of live-time 
for 1997 and 1998 combined. The peak of the distribution is very well fit by a
Gaussian distribution.
Assuming that the dark noise peak is Gaussian and that the data from the 
different OMs are uncorrelated, we sould expect one entry per year above a value of
$1235$ Hz and 1 entry per 215 days for $\res>1200$ Hz.
\par
In order to get rid of more background events with large $RES$,
we  need to develop new tools using the more detailed data information at our disposal. 
We can distinguish between three classes of events in the experimental data:
\begin{enumerate}
\item pure dark noise background,
\item supernova signal + pure dark noise (i.e. our signal events),
\item events where instrumental noise is present.
\end{enumerate}
We will start by assuming that the dark noise, as well as
the excess counts caused by the signal, are Gaussian and that all OMs are
equivalent.
If a supernova would be responsible for an
event with high $RES$, we would expect the noise
rates of all OMs to increase coherently. 
The first and second classes can easily be separated by cutting on the
value of the summed noise residual $RES$.
The third class is more problematic, since those events can yield a large
$RES$-value due to an increase in noise rates caused by disturbances occurring
outside the detector itself.
In that case, it is reasonable to suspect that the parts of the front-end detector
which are affected are located in the control room at the surface of the ice and 
that different components
of the electronics pick up noise in different ways. If such events are caused by
some unstable channels, rather than by an overall increase of the noise in
the detector, this should be a clear signature for a background event. 
\par
With the aim to remove these fake supernova candidates, we consider the likelihood of each
event. Furthermore, since the rates are nearly Gaussian,  a $\chi^2$ function can
be used instead of a likelihood function:
\begin{eqnarray}
\label{eq:chi2}
\chi^2=\sum_{i=1}^{N_\mathrm{OM}}\left(\frac{n_i-\mu_i-\Delta\mu} 
{\sigma_i}\right)^2
\end{eqnarray}
where $n_i$ is the measured noise of each of the $N_\mathrm{OM}$ optical modules and $\mu_i$
is its expectation value --the moving average of the $i$:th OM will be used as an approximation.\par
The difference ($n_i-\mu_i$) is the deviation of the noise from its
moving average. $\Delta\mu$ represents the expected number of counts per OM per sec,
which are  caused by  the signal, and comes in addition to the mean dark noise. Typically 100 counts in 10
sec, or 10 Hz per OM would be caused by a SN1987A-type supernova at the center of the
Milky Way.  Finally, $\sigma_i$ is the standard deviation for OM number ($i$).
The corresponding variance is $\sigma_i^2=\sigma_{\mathrm{OM}_i}^2+\deltamu/{\Delta t}$, where 
the term $\deltamu/{\Delta t}$ is added quadratically to the spread of the OM dark noise 
$\sigma_{\mathrm{OM}_i}$ to account for  fluctuations in the signal, keeping in mind that  \deltamu~is a rate
measured over ${\Delta t}=10$ second but given in [Hz]. For each analyzed run, a
Gaussian fit is made of the noise rate of OM number $i$ after the moving-average
correction and the resulting standard deviation $\sigma_{\mathrm{OM}_i}$ is used.\\
\par
\begin{figure}[htp]
 \centering
  \includegraphics*[width=0.45\textwidth]{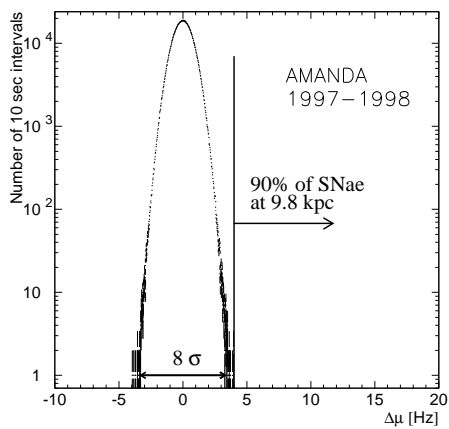}
        \caption{
\footnotesize{ Distribution of $\Delta\mu$ after application
of cuts, for 215 days of live-time.
The vertical line is set at the level where we get a rate of one background event per
year. 90$\%$ of supernova neutrino bursts located at 9.8 kpc distance would be seen above
that cut. 
}}
        \label{fig:after_liky}
\end{figure}

 Note that the same $\chi^2$ function as in Eq.\ref{eq:chi2} would be set up in the hypothetical
situation where one would measure a standard supernova-signal $N_\mathrm{OM}$ times using
only one OM.
In the case of classes (1) and (2) of events with homogeneous noise, the $\chi^2$-value 
will be small.
By contrast, that value is expected to be large for any 
$\Delta\mu$ if the noise recorded by the OMs is not uniform
throughout the detector. By cutting on $\chi^2$, one can thus expect to reject
events of class (3) above.\\
Near the background, \ie when $\deltamu/{\Delta t}<<\sigma_{\mathrm{OM}_i}^2$,  we can solve 
for $\Delta\mu$, by minimizing Eq.~\ref{eq:chi2} w.r.t. \deltamu:
\begin{eqnarray}
\label{eq:deltamu}
\lefteqn{ \Delta\mu=\frac{1} {\sum_{i=1}^{N_\mathrm{OM}}
   {1/\sigma_i^2}} \times} \nonumber \\ &               & \hspace{2cm}\sum_{i=1}^{N_\mathrm{OM}} 
\frac{1} {\sigma_i}\left(\frac{n_i-\mu_i} {\sigma_i}\right)
\end{eqnarray}
so that we get an actual measurement of the supernova-induced rate excess per 
OM, \deltamu, predicted in Eq.~\ref{eq:halzen_signal}. With the assumption that all OMs 
are equivalent, the variable $RES$ is the product $N_\mathrm{OM} \times \Delta\mu$.
The standard deviation of \deltamu~is:
\begin{eqnarray}
\label{eq:stddev}
\sigma_{\Delta_\mu}^{\mathrm {noise}} = \sqrt{\frac{1} {\sum_{i=1}^{N_\mathrm{OM}}{1/\sigma_i^2}}}
\end{eqnarray}
This expression for $\sigma_{\Delta_\mu}^{\mathrm {noise}}$ yields the standard deviation of
the background distribution, {\it i.e.} when $\Delta\mu=0$ (see Fig.\ref{fig:after_liky}).
The characteristics of the individual OMs are taken into
account in the computation of $\Delta\mu$ with the different values of ${\sigma_i}$.\\
\par We can also introduce a module-dependent sensitivity  in Eq.\ref{eq:chi2} to treat
properly the variation of the absorption-length of the ice as a function of
depth \cite{icrc-kurt}, modifying Eqs.\ref{eq:chi2} and \ref{eq:stddev}:
\begin{eqnarray}
\label{eq:chi2_eps}
\chi^2=\sum_{i=1}^{N_\mathrm{OM}}\left(\frac{n_i-\mu_i-\epsilon_i\cdot\Delta\mu} {\sigma_i}\right)^2
\end{eqnarray}
\begin{eqnarray}
\label{eq:stddev_eps}
\sigma_{\Delta_\mu}^{\mathrm {noise}} = \sqrt{\frac{1} {\sum_{i=1}^{N_\mathrm{OM}}{(\epsilon_i/\sigma_i)^2}}}
\end{eqnarray}
Here $\epsilon_i$ is the relative sensitivity of ${\mathrm OM}_i$, calculated as
$V^i_\mathrm{eff}/V^0_\mathrm{eff}$ where $V^0_\mathrm{eff}=414\: {\mathrm m^3}$ is the
reference volume mentioned is section~\ref{sec:lownue}.\par

\begin{figure}[htp]
 \centering
  \includegraphics*[width=0.45\textwidth]{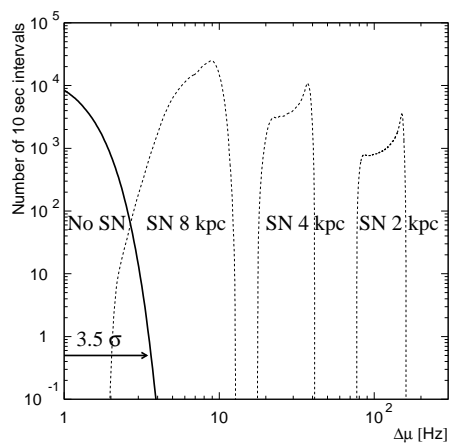}
        \caption{
\footnotesize{ Background distribution and expected 
SN signal of $\Delta\mu$ for supernovae at different distances,
in kpc. The smearing of the signal
due to the fixed 10 sec observation window is taken into account (see Fig.\ref{fig:sn_onset}).
Notice the logarithmic scale in $\Delta\mu$.
}}
        \label{fig:more_sne}
\end{figure}
Fig.\ref{fig:more_sne} shows the distribution of the $\Delta\mu$ signal expectation
for supernovae of type SN1987-A located at different distances,  including the effect of the
fixed 10 sec time-window location. The expected distribution in the absence of a supernova
is also shown, for comparison.\par 
The resulting $\Delta\mu$ distribution for the combined 1997 and 1998 data-sets can be seen in 
Fig.\ref{fig:after_liky}, where a total live-time of 215 days has been accumulated. 
A cut on $\chi^{2}/\mathrm{n.d.f.}<1.3$ has been applied (leaving 99.9$\%$ of events), removing 
fake events which could still be seen in Fig.\ref{fig:only_ma}.
Without the $\chi^{2}/\mathrm{n.d.f.}$ cut, but cutting on $\deltamu > 4.0$ Hz, we 
have 1.7 background events per week.
The spread of the background $\sigma_{\Delta\mu}^{\mathrm {noise}}=0.8$ Hz is the same for both years.
This reflects the fact that the eleven additional OMs used in the 1998 set have high
noise levels, and  their spreads contribute little to reducing the total spread in
Eq.\ref{eq:stddev_eps}.
\par
Considering the background distribution of
Fig.\ref{fig:after_liky}, a SN signal detected with
$90\%$ efficiency at a distance of 8, 4 and 2 kpc would correspond to an excess of 7,
29 and 116 standard deviations (see Fig.\ref{fig:more_sne}).
The line on Fig.\ref{fig:after_liky} indicates the \deltamu-level ($=4$ Hz) above which we
would expect one {\it fake event} per year. Since the distribution in the figure is
well fitted with a Gaussian, this number is found by fitting the data and integrating from that point
and up to infinity. A cut on $\Delta\mu \ge 4$ Hz corresponds to a $90\%$ efficiency for a SN
1987A-like supernova  located at a distance of 9.8 kpc. 
\section{Supernova sensitivity}
\label{sec:performance}
The results of the analysis, along with numerical
simulations, can be used to further assess the performance of \amanda as a
supernova-burst detector.\par
 Since the fitted  noise-rate excess per OM is centered
at zero in the absence of any supernova and has a known Gaussian spread,  the number of
dark noise background events  is  easily calculated for a given cut on \deltamu. \par
The signal that we 
expect to see is characterized by a frequency and a magnitude, both of which must be
estimated to decide how to compute our signal-to-noise.
 The more elaborate estimates of the frequency of gravitational collapses are made using 
various methods and combining
them \cite{sn_rate}. We will limit ourselves to the most conservative guess of 1
supernova/century given in \cite{suzuki}, since the performance of the detector does 
not depend strongly on that number, as will be shown.\par
The size of the signal depends on the
distribution of progenitor stars in the Galaxy \cite{piran} and on the
supernova neutrino-luminosity. The fraction of stars which could undergo a
gravitational collapse is shown in 
Fig.\ref{fig:fraction}. We decided to use the measured luminosity of SN 1987A as
an estimate of the signal-strength, since it is so far the only observation of an actual
event.\par
\begin{figure}[h]
  \centering
  \includegraphics*[width=0.45\textwidth]{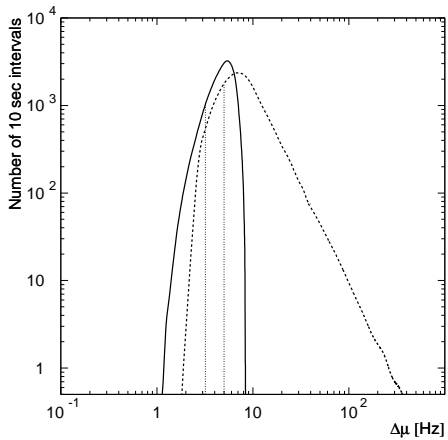}
        \caption{
\footnotesize{Expected $\Delta\mu$ distributions {\it at} (full curve) and {\it within} (dotted curve) a
distance of 10 kpc, shown in log-log scale. The dominant contribution to the signal is
 given by the closest stars, which appear in the tail of the distributions at high
$\Delta\mu$ values. 
The vertical lines mark the $90\%$ detection efficiency lower limit.
}}
  \label{fig:delta_signal}
\end{figure}
Simulation studies were made in order to estimate the signal, for which we used the
time-profile described in section \ref{sec:theory}.
The effect of the 10 sec time-window  inefficiency  (see Sec.\ref{sec:data}) , as well as the Gaussian
smearing due to dark-noise fluctuations, was then applied to a signal of strength given by 
Eq.~\ref{eq:halzen_signal}.
As an example, the resulting \deltamu -distribution for a supernova occurring at a distance of 
10 kpc can be seen in Fig.\ref{fig:delta_signal}(full curve). 
If we fold in the probability distribution shown in Fig.\ref{fig:fraction}, 
we can also obtain the signal distribution for supernovae occurring {\it within},
rather then {\it at} a radius of 10 kpc from the Earth (see Fig.\ref{fig:delta_signal}, dotted curve).
\par
\begin{figure}[h]
  \centering
  \includegraphics*[width=0.45\textwidth]{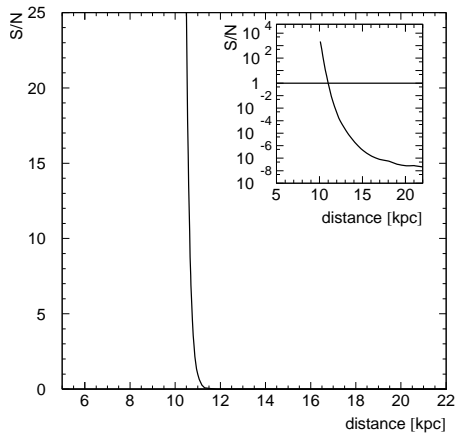}
        \caption{
\footnotesize{ The $S/N$-ratio as a function of distance for strings 1-10
($S/N > 1$ within 11 kpc, i.e. 60\% of the Galaxy). $S/N= 1$ corresponds to a cut at
$\Delta\mu = 5.5$ Hz. Inset: the same curve seen in a logarithmic scale.
}}
  \label{fig:signal_noise}
\end{figure}

With these definitions of background and signal ({\it i.e.}
everything within a sphere of a given radius), it is now possible to define a
performance parameter which we call signal-to-noise $S/N$ at a given distance.
For this purpose, the \deltamu -distribution of progenitor stars within that distance is
computed as well as the lower cut
which keeps $90\%$ of stars above it.
The signal $S$ above the \deltamu -cut is calculated, 
taking all efficiencies into account, as well as the estimated rate.
Finally, the noise $N$ from the dark noise background for the same period of time and
same cut on \deltamu~is readily computed from the Gaussian assumptions.
Fig.\ref{fig:signal_noise} shows the resulting $S/N$ parameter as a function of 
distance from the Earth.
At a characteristic radius, a drastic drop in $S/N$ from very large values down to essentially
zero occurs, which characterizes the detector performance.
%

\begin{figure}[h]
  \centering
  \includegraphics*[width=0.45\textwidth]{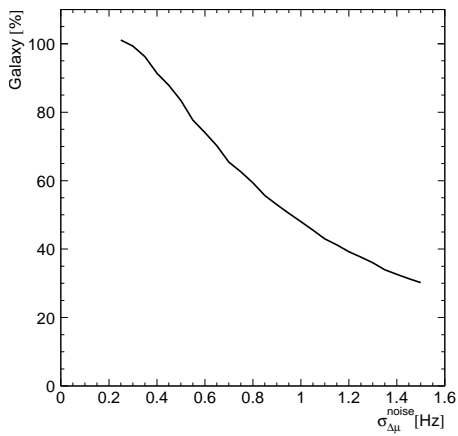}
        \caption{
\footnotesize{Percentage of the Galaxy coverage
versus $\sigma_{\Delta\mu}^{\mathrm {noise}}$.
Note that no specific detector setup is assumed in this figure.
}}
  \label{fig:galaxy}
\end{figure}

We will therefore take the point at which $S/N=1$ as a natural measure of the 
performance of different detector configurations. Note that the distance that
corresponds to that point does not depend strongly on the chosen $S/N$ value, due to
the sharpness of the drop.
For instance, a larger estimate for the gravitational collapse rate would not change
that characteristic distance significantly.
Indeed, using the different estimates found in the literature,
of 1 SN every 100~\cite{suzuki}, 47~\cite{sn_rate} and 11~\cite{bahcall} years,
the $S/N=1$ point would be located at 11 ($\sim 60\%$ Galaxy coverage), 11.3($63\%$)
and 11.8($66\%$) kpc, respectively.
The small differences are due to the steepness of the curve (see Fig.\ref{fig:signal_noise}).
\amanda B10 is thus able to detect supernovae within a distance of 11 kpc with a S/N
$> 1$, {\it i.e.} to cover $\sim60\%$ of the stars in the Milky Way with one background
fake per century. The coverage for detector configurations with other
$\sigma_{\Delta\mu}^{\mathrm {noise}}$-values is shown in Fig.\ref{fig:galaxy}.
If one lowers the requirement to accepting one background event per year and a
supernova detection probability of $90\%$, the Galactic
coverage is $70\%$. Even though the additional OMs in strings 5-10 have worse noise
characteristics, they provide a significant improvement. Observe however, that this is mainly due to
the very steep behavior of the distribution in  Fig.\ref{fig:fraction} in the range of
7-14 kpc around the bulge, where the star density is the highest.\par
The non-observation of any supernova signal can be used to put an upper limit on the rate of
gravitational stellar collapses:
\begin{eqnarray}
\label{eq:upper-limit}
R_{\mathrm{c.l.}} =  \frac{n_{\mathrm{c.l.}}}{\eta \epsilon \times T}
\end{eqnarray}
where $n_{\mathrm{c.l.}}$ is the upper bound at the 90$\%$ c.l. on the expected number of
events, $\eta$ is the Galactic coverage (70$\%$), $\epsilon$ is the detection
efficiency ($90\%$) and $T$ is the live time (215 days).
Using the procedure described in \cite{feldman} (with no candidates observed and one
estimated background fake per year), we obtain an upper limit on the rate of
gravitational stellar collapses at the $90\%$ c.l., of 4.3 events per year.
\par

\section{Amanda Supernova Trigger Algorithm (\asta)}
\label{sec:trigger}
The goal of the trigger is to detect a supernova event in real
time, in order to send out a prompt alarm to the astronomical
community, ahead of its optical observation.
A future implementation of \asta~ at the South Pole will make it possible to join
SNEWS~\cite{scholberg2}, a supernova early warning network of 
neutrino detectors.
A first version of \asta~ based on 11 days of data was
presented in \cite{silves}.\par
The online algorithm must be able to suppress background events
which could fake a SN neutrino burst, especially those due to non-Gaussian variations of the noise
rates.
We must use a moving average which uses exclusively events occurring to the left, 
\ie before the considered
time-bin.  We will refer to it as $MA_\mathrm{left}$, as opposed to $MA_\mathrm{sym}$, with 
the difference that the sum in Eq.\ref{eq:moving_average} is now taken from $(i-\tw)$ to $i$.
In order to accommodate for faster variations of the data, we define an additional 
variable, the derivative of 
$MA_\mathrm{left}$ for each event $(i)$:
 \begin{eqnarray}
\label{eq:der}
\lefteqn{DER(i)=}\\ && [MA_\mathrm{left}(i)-MA_\mathrm{left}(i-1)]/{\Delta t}\nonumber
\end{eqnarray}
where $\Delta t=100$ sec.
It has been tested with numerical simulations that during normal data-taking
conditions,  the occurrence of  a SN signal should not increase $DER(i)$
significantly. 
\par
Using the same OM selection as in the offline-analysis was found to be unsuitable for
\asta, rejecting a large percentage of the events after applying a cut on $\chi^2/\mathrm{n.d.f.}<1.3$.
Instead, we removed 117 known unstable OMs and then applied the cleaning algorithm
described in section~\ref{sec:data} on the 1998 data sample. This resulted in a set of 170 OMs
that were then used to test the cuts on {\it both} the 1997 and 1998 data. 
 \begin{figure}[htp]
  \centering
   \includegraphics*[width=0.5\textwidth]{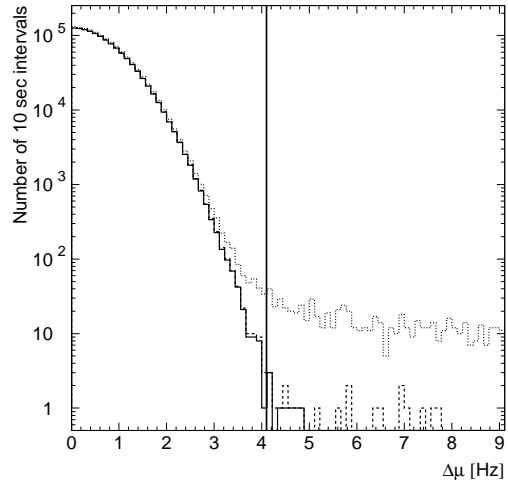}
            \caption{
 \footnotesize {Tail of the background distribution after different cut levels:
(dotted line) 'no cuts', (dashed line) '$\chi^2+DER$ cut', (solid line) 'all
cuts'. The vertical line indicates the $\Delta\mu$-cut that leaves one
statistical fake per year. The live time is 296 days.
 }}
            \label{fig:levels}
 \end{figure}
Fast time variations of the data ($S_\mathrm{tot}$) are not always
corrected fully by moving-average subtraction.
Taking this into account we decided to iterate, \ie
apply the moving-average correction a second time, on $RES$:
\begin{eqnarray}
\label{eq:resres}
\lefteqn{RES_\mathrm{iter}(RES(i)) = }\nonumber\\
&&\hspace{0.5cm}RES(i) - MA_\mathrm{left}(RES(i)) 
\end{eqnarray}
A new $\Delta\mu$ can be calculated, as well as new variables $\chi^2_\mathrm{iter}$ 
and $DER_\mathrm{iter}$.
The final cuts applied are: $\Delta\mu > 4.1$ Hz, $|DER| < 1 \, \mathrm{s^{-2}}$, 
$\chi^2/\mathrm{n.d.f.} < 1.3$, $|DER_\mathrm{iter}| < 1 \, \mathrm{s^{-2}}$ and 
$\chi^2_\mathrm{iter}/\mathrm{n.d.f.} < 1.3$,
leaving five
triggers only, \ie one every two months (see Fig.\ref{fig:levels}).
Note that these requirements leave the efficiency to trigger on a SN during stable data-taking
conditions unaffected.
In order to estimate the performance of the cuts as previously, we use  $\sigma_{\Delta\mu}^{\mathrm {noise}} = 0.85$
Hz and the cut $\Delta\mu>4.1$ Hz, resulting in a $65\%$ coverage of the
Galaxy with $90\%$ detection efficiency. On top of the five fakes found experimentally, we expect
statistically one background event per year.
We note that all fakes have low significance 
and that already a cut on
$\Delta\mu>5$ Hz can get rid of them, which would lower the Galactic coverage to $55\%$. 

\section{Discussion}
\label{sec:discussion}
The good overall performance of the detector is mainly due to the low noise of the OMs
in strings 1-4. However, even if all OMs would be of the same type as those, 800 of them would
still be required to cover $100\%$ of the Milky Way. From Eq.\ref{eq:stddev_eps} and
Fig.\ref{fig:galaxy}, it
is clear that $\sigma_{\Delta\mu}^{\mathrm {noise}}$ is a crucial parameter for the detector performance
and must be kept as small as possible.\par
Assuming a uniform detector with only one type of OMs, Eq.\ref{eq:stddev_eps} becomes:
\begin{eqnarray}
\label{eq:discussion2}
\sigma_{\Delta_\mu}^{\mathrm {noise}} = \frac {\sigma_\mathrm{OM}}{\epsilon\sqrt{N_\mathrm{OM}}}
\end{eqnarray}
One sees from Eq.\ref{eq:discussion2} that increasing $N_\mathrm{OM}$ is a less efficient
way of reducing $\sigma_{\Delta\mu}^{\mathrm {noise}}$ than increasing the collection efficiency
$\epsilon$ or reducing $\sigma_\mathrm{OM}$, the spread in the dark-noise rate of each OM.
Several technical improvements to the OMs can be considered, which affect
$\sigma_\mathrm{OM}$, $\epsilon$ and $\sigma_{\Delta\mu}^{\mathrm {noise}}$:
\begin{itemize}
\item coating the OMs with wavelength-shifters would increase the photon-collection
efficiency without affecting the noise rates.
\item using PMTs with a larger cathode area would increase the efficiency, even though
the noise level would go up.
\item reducing the potassium levels in the pressure glass would reduce the dark noise spread.
\item Fig.\ref{fig:veff} shows that there are ice-depths with larger transparency which are most
suitable for SN detection. In fact, the transparency is expected 
to be extremely good at shallower depths \cite{science_paper,indirect_lambda}. However, the muon flux
is larger closer to the ice surface, producing more background Cherenkov light.
\end{itemize}
\vfill{\ }  
All these improvements have been investigated and we estimate that they can reduce the
value of $\sigma_{\Delta\mu}^{\mathrm {noise}}$ by $20-40\%$ each.
\par
Detailed investigations on after-pulsing in the range up to 1 msec are currently being 
pursued in order to optimize the choice of artificial dead-time. The effects of raising
the PMT thresholds remain to be investigated.
Finally, it may be noted that a sliding time-window, fitted to the supernova
time-profile, rather than applied at arbitrary times, would increase the galactic
coverage of the detector studied in this paper by an estimated $11\%$. One could even
optimize the size of the window to fit
the peaked part of the profile so that the ratio $\epsilon / \sigma_\mathrm{OM}$ is
maximized. An additional $10\%$ could be gained in this way, even though it is clear that
such optimization is partly model-dependent (for the model used here, a 4 sec time window would be optimal).
\par
An uncertainty in the strength of the signal is the \antinue ~temperature. In this paper, we use
the same assumption as in \cite{sn_halzen} ($T_{\antinue} = 4$ MeV), but there is no precise 
prediction for this value and it may be lower (see \eg \cite{jegerlehner}). Furthermore, corrections 
to inverse beta-decay cross-sections due to recoil and weak magnetism \cite{beacom_corrections} 
would lower the resulting positron energy somewhat.
On the other hand, this can be compensated in some measure by the performance of the present detector (680 OMs), 
which is expected to be higher than the configuration studied in this paper.
With the planned transition to a detector of kilometer-cubed size, using thousands of OMs, 
full coverage of the Galaxy will be attained. 
\par
One of the greatest assets of \amanda with its low background rate is that it has the potential 
to see a supernova independently from other detectors. In the event of a discovery, the time 
profile and intensity of the burst could be studied. These aspects will be investigated in the future.
Should an event occur in our Galaxy, the information that could presently be provided to the rest of the 
supernova community, is an estimate of $\Delta\mu$ and a time stamp associated with the event.

\section{Conclusions}
\label{sec:conclusions}
\amanda primarily aims at detecting high-energy neutrinos. However, by monitoring
bursts of low-energy neutrinos, it is operating as a gravitational collapse detector as 
well.
In the course of this analysis, a method has been developed and applied to 
filter the data and remove unstable periods and OMs.
from disturbances.
Furthermore, the $\chi^2$-based analysis method used here takes into account
the varying statistical properties of different OMs.
The sets of data taken in 1997 and 1998, corresponding to 215 days of live-time after
cleaning, have been fully searched for supernova bursts occurring in our galaxy.
No significant candidates were found.
A conservative model of the supernovae-distribution in space and time has been used to
assess the performance of the detector, resulting in an estimated coverage of the
galaxy of $ 70 \%$ with $90\%$ efficiency, with one expected background fake per
year. This performance is satisfying, the more so since the primary design of \amanda is for 
high-energy neutrino detection.
A real-time supernova detection algorithm, has been developed and tested on the 296
days of live time available. The algorithm turned out to be sufficiently robust and
resulted in five fakes with the set of cuts chosen. We find that 800 OMs of the same type
as those used in the first four strings of the array would suffice to cover the entire
galaxy. It is also shown that this number can be reduced significantly, either by
reducing the spread of the dark noise, or by improving the collection sensitivity of the OMs.

\section*{Acknowledgments}
We are grateful to J. Beacom for discussions concerning the predicted signal and 
for very helpful comments on the manuscript.\par
This research was supported by the U.S. NSF office of Polar Programs
and Physics Division, the U. of Wisconsin Alumni Research Foundation,
the U.S. DoE, the Swedish Natural Science Research Council, the
Swedish Polar Research Secretariat, the Knut and Alice Wallenberg
Foundation, Sweden, the German Ministry for Education and Research,
the US National Energy Research Scientific Computing Center (supported
by the U.S. DoE), U.C.-Irvine AENEAS Supercomputer Facility, and
Deutsche Forschungsgemeinschaft (DFG).  D.F.C. acknowledges the
support of the NSF CAREER program.  P. Desiati was supported by the
Koerber Foundation (Germany).  C.P.H. received support from the EU 4th
framework of Training and Mobility of Researchers.  St. H. is
supported by the DFG (Germany).  P. Loaiza was supported by a grant
from the Swedish STINT program.

\baselineskip 10pt
\begin{bibliography}{sn-paper}
\end{bibliography}


\begin{thebibliography}{10}
\expandafter\ifx\csname url\endcsname\relax
  \def\url#1{\texttt{#1}}\fi
\expandafter\ifx\csname urlprefix\endcsname\relax\def\urlprefix{URL }\fi

\bibitem{IMB}
R.~M. {{Bionta} et al.}, \prl {\bf 58} (1987) 1494--1496.

\bibitem{k2}
K.~Hirata, et~al., \prl {\bf58} (1987) 1490--1493.

\bibitem{hirata88}
K.~Hirata, et~al., \prd {\bf38} (1988) 448--458.

\bibitem{scholberg3}
K.~{Scholberg}, in: {\it 19th International Conference on Neutrino Physics and
  Astrophysics}, 2000, hep-ex/0008044.

\bibitem{baksan}
E.~Alexeyev, et~al., \npbsuppl {\bf35} (1994) 270--272.

\bibitem{b4}
E.~Andres, et~al., \app {\bf13} (2000) 1--20.

\bibitem{baikal99}
V.~A. Balkanov, et~al., Phys. Atom. Nucl. {\bf63} (2000) 951.

\bibitem{antares2000}
C.~Carloganu, Nucl. Phys. Proc. Suppl. {\bf85} (2000) 146--152.

\bibitem{nestor2000}
S.~Bottai, et~al., Nucl. Phys. Proc. Suppl. {\bf85} (2000) 153--156.

\bibitem{nemo_icrc99}
T.~Montaruli, et~al., {\it Proceedings of the 26th International Cosmic Ray
  Conference (ICRC 99), Salt Lake City, UT} HE.6.3.06.

\bibitem{sn_halzen}
F.~Halzen, J.~E. Jacobsen, E.~Zas, \prd {\bf 49} (1994) 1758--1761.

\bibitem{pryor}
C.~Pryor, C.~Roos, M.~Webster, \apj {\bf329} (1988) 335--338.

\bibitem{sn_rome}
R.~Wischnewski, et~al., in: {\it Proceedings of the 24th International Cosmic
  Ray Conference, Rome}, Vol.~{\bf 1}, Italy, 1995, pp. 658--661.

\bibitem{desy_proposal}
A.Biron, et~al., {\it Proposal submitted to the Physics Research Committee,
  DESY}, {PRC} 97/05 (1997).

\bibitem{icrc99}
R.~{{Wischnewski} et al.}, {\it Proceedings of the 26th International Cosmic
  Ray Conference (ICRC 99), Salt Lake City, UT} HE.4.2.07.

\bibitem{kennicutt}
R.~C. {Kennicutt}, \apj {\bf 277} (1984) 361.

\bibitem{burrows}
A.~Burrows, D.~Klein, R.~Gandhi, \prd {\bf 45} (1992) 3361--3385.

\bibitem{raffelt}
G.~Raffelt, {\it Stars as laboratories for fundamental physics}, {The
  University of Chicago Press}, 1996.

\bibitem{beacomsuperk}
J.~F. {Beacom}, P.~{Vogel}, \prd {\bf58} (1998) 053010.

\bibitem{beacom}
J.~F. {Beacom}, P.~{Vogel}, \prd {\bf60} (1999) 033007.

\bibitem{soneira}
J.~N. {Bahcall}, R.~M. {Soneira}, \apj {\bf 44} (1980) 73.

\bibitem{piran}
J.~N. {Bahcall}, T.~{Piran}, \apj {\bf 267} (1983) L77.

\bibitem{sn_rate}
G.~A. Tammann, W.~{L\" offler}, A.~{Schr\" oder}, \apjs {\bf 92} (1994)
  487--493.

\bibitem{superkpro}
Y.~{Fukuda}, in {\it First International Workshop on the Supernova Early Alert
  Network}, Boston University, unpublished, 1998.

\bibitem{lvd}
{{The LVD} Collaboration}, in: {\it Proceedings of the 26th International
  Cosmic Ray Conference (ICRC 99), Salt Lake City, UT}, 1999, {HE.4.2.08}.

\bibitem{halzen92}
F.~Halzen, T.~Stanev, E.~Zas, \prd {\bf45} (1992) 362--376.

\bibitem{halzennew}
F.~{Halzen}, J.~E. {Jacobsen}, E.~{Zas}, \prd {\bf 53} (1996) 7359.

\bibitem{jacobsen}
J.~E. Jacobsen, Ph.D. thesis, University of Wisconsin, Madison (1996).

\bibitem{icrc-kurt}
K.~{{Woschnagg} et al.}, Proceedings of the 26th International Cosmic Ray
  Conference (ICRC 99), Salt Lake City, UT HE.4.1.15.

\bibitem{teich1}
B.~E.~A. Saleh, J.~T. Tavolacci, M.~C. Teich, {IEEE} J. Quant. Electron. {\bf
  QE-17} (1981) 2341--2350.

\bibitem{brockwell}
P.~J. {Brockwell}, R.~A. {Davis}, {\it Introduction to Time Series and
  Forecasting}, Springer Verlag, 1996.

\bibitem{suzuki}
A.~{Suzuki}, {\it Supernova Neutrinos, in Physics and Astrophysics of
  Neutrinos}, Springer Verlag, 1994.

\bibitem{bahcall}
J.~N. {Bahcall}, {{\it Neutrino Astrophysics}}, Cambridge University Press,
  1989.

\bibitem{feldman}
G.~Feldman, R.~Cousins, \prd {\bf57} (1998) 3873--3889.

\bibitem{scholberg2}
K.~{Scholberg}, in: {\it Proceedings of the 3rd Amaldi Conference on
  Gravitational Waves}, 1999, astro-ph/9911359.

\bibitem{silves}
A.~Silvestri, Diploma thesis: DESY-THESIS-2000-028, Zeuthen, {ISSN} 1435-8085
  (2000).

\bibitem{science_paper}
P.Askebjer, et~al., Science {\bf 267} (1995) 1147.

\bibitem{indirect_lambda}
S.Tilav, et~al., in: {\it Proceedings of the 24th International Cosmic Ray
  Conference, Rome}, Vol.~{\bf 1}, 1995, p. 1011.

\bibitem{jegerlehner}
B.~Jegerlehner, F.~Neubig, G.~Raffelt, \prd {\bf54} (1996) 1194--1203.

\bibitem{beacom_corrections}
P.~Vogel, J.~Beacom, \prd {\bf60} (1999) 053003.

\end{thebibliography}
\end{document}